\pdfoutput=1 
\documentclass[aip, jcp, amsmath,amssymb,floatfix,reprint]{revtex4-1}

\usepackage[utf8]{inputenc}
\usepackage{graphicx}  
\usepackage{natmove}

\usepackage[hidelinks]{hyperref}

\usepackage{pdfpages}
\usepackage{pgffor}
\makeatletter
\AtBeginDocument{\let\LS@rot\@undefined}
\makeatother
\def\supplementfilename{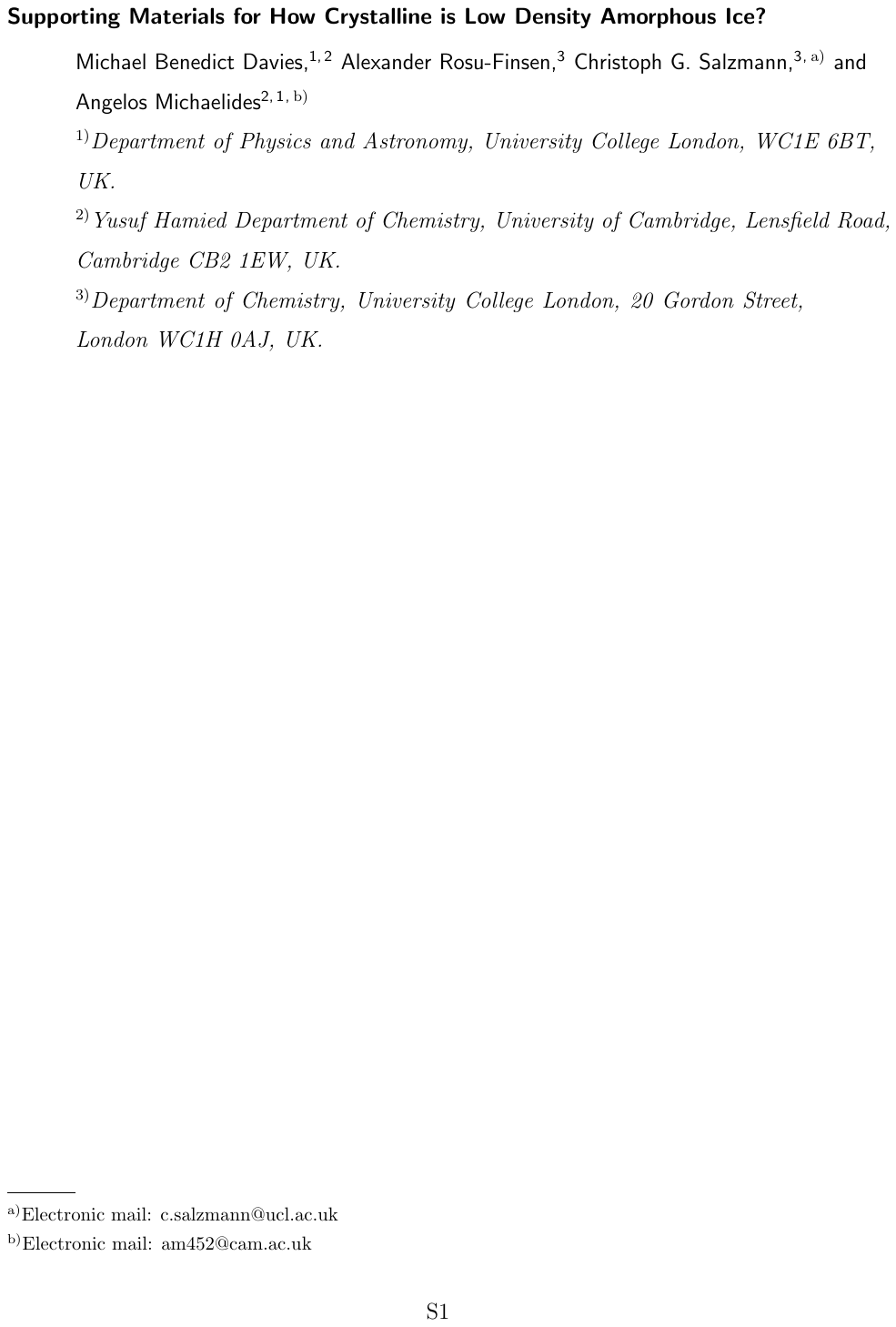}
\pdfximage{\supplementfilename}
\def\numbersupplementpages{\the\pdflastximagepages}

\begin{document}
\title{How Crystalline is Low-Density Amorphous Ice?}

\author{Michael Benedict Davies}
\affiliation{Department of Physics and Astronomy, University College London, WC1E 6BT, UK.}
\affiliation{Yusuf Hamied Department of Chemistry, University of Cambridge, Lensfield Road, Cambridge CB2 1EW, UK.}
\author{Alexander Rosu-Finsen}
\affiliation{Department of Chemistry, University College London, 20 Gordon Street, London WC1H 0AJ, UK.}
\author{Christoph G. Salzmann}
\email{c.salzmann@ucl.ac.uk}
\affiliation{Department of Chemistry, University College London, 20 Gordon Street, London WC1H 0AJ, UK.}
\author{Angelos Michaelides}
\email{am452@cam.ac.uk}
\affiliation{Yusuf Hamied Department of Chemistry, University of Cambridge, Lensfield Road, Cambridge CB2 1EW, UK.}
\affiliation{Department of Physics and Astronomy, University College London, WC1E 6BT, UK.}

\begin{abstract}
Low-density amorphous ice (LDA) is one of the most common solid materials in the Universe and a key material for understanding the many famous anomalies of liquid water.
Yet, despite its significance and its discovery dating nearly 90 years, the structure of LDA is debated.
It is unclear if LDA is a glassy state representing a liquid or a heavily disordered crystal; indeed, two forms (LDA-I and LDA-II) have been discussed as amorphous and partially crystalline in the literature, respectively.
Here, with two widely used water models, we show that the experimental structure factor of LDA is best reproduced computationally by a partially crystalline structure.
Models for both LDA-I and LDA-II are highly similar, with differences only due to subtle differences in crystallinity and/or experimental error.
Further support for this structural model of LDA comes from experiment:
if LDA is partially crystalline, then its route to formation should result in different nanocrystallite cubicities, and thus give rise to different cubicities upon recrystallisation. This memory effect of LDA's creation route is observed and it is incompatible with a fully amorphous material.
The results present a unified computational and experimental view that LDA is not fully amorphous but instead a partially crystalline material.
This impacts LDA's many roles in nature and potentially our understanding of liquid water.
Furthermore, the ``re-identification'' of such an intensely studied material highlights that great care will be needed when classifying the nature of glassy materials going forward.
\end{abstract}

\maketitle

\date{\today}


\section{Introduction}
The matrix of life, the universal solvent, a powerful corrosive, and the strangest fluid -- these are just some of the names for perhaps the most fascinating and important substance in nature: water.
It is the many anomalies of water -- over 70 have been compiled \cite{LSBUwebsite} -- that facilitate its multitude of names and moreover shape the face of the Earth.
Yet, despite intensive research efforts dating back to antiquity, a complete understanding of water remains a challenge.
There are many physical states that water can exist in; for instance, 20 polymorphs of ice have now been created in the laboratory \cite{Millot2019, SalzIceXIX, Yamane2021, Gasser2021, Prakapenka2021, Salz2019advances}.
Of all water's forms, the most abundant in the Universe is found in the dense molecular clouds from which stars and planets are born, and as the bulk matter in comets: low-density amorphous ice (LDA) \cite{JenniskensBlake1994, Burke2010}.
LDA is of great scientific interest, in part due to its involvement in cosmological phenomena (including a possible role in the origin of life \cite{Ehrenfreund2000}), but also due to its potential key role in explaining water's anomalies.
Specifically, whether LDA has a corresponding liquid state is a cornerstone of the heavily debated two-state liquid model, which could account for many of water's thermodynamic anomalies \cite{Poole1992Nature, mishima1998decompression, Tse1999, Schober2000, Debenedetti2003SupercooledReview, Geil2004rsc, Limmer2014Theory, Shephard2017derailed, Debenedetti2020Science, tulk2019absence, Kim2020Science}.

The name of LDA originates from the fact that its density of 0.94 g cm$^{-3}$ is lower than the 1 g cm$^{-3}$ of liquid water, and because it appears to be an amorphous solid according to diffraction studies \cite{Mishima1984Nature}.
However, despite the importance of LDA, there is ambiguity with respect to its structural nature with some studies pointing towards a more ``crystal-like'' nature than a purely glassy state \cite{Li1997inelastic, Tse1999, Schober2000, Andersson2002thermal, Geil2004rsc, Andersson2005, Andersson2007dielectric, Tse2008xray, Elsaesser2010}. %
Multiple experimental avenues for creating LDA have been discovered 
\cite{Burton1935, Burton1935Nature, JenniskensBlake1994,Bruggeller1980, Mayer1982, Mayer1985,Bruggeller1980, Mayer1982, Mayer1985,Mishima1984Nature, Mishima1985Nature, Kouchi1990, Baratta1991, Leto2003, Klug1989, Klotz1999}. 
And it is identified on the basis of broad features in diffraction.
However, discerning the exact local structural nature of LDA from diffraction patterns is problematic. 
Amann-Winkel et al. distinguished three scenarios which are of interest for diffraction patterns of amorphous ice: a purely amorphous material, grains of crystalline ice embedded in an amorphous matrix, and a completely polycrystalline material \cite{Amann2016}. 
The structural nature of LDA is further complicated by subtle differences that can be exhibited by samples depending on their preparation route (for example, LDA-I, LDA-II and LDA(ice VIII) \cite{Winkel2009LDA2, shephard2016LDAVIII}).
Using neutron and X-ray diffraction respectively, Finney et al. and Mariedahl et al. found the first coordination shell of LDA is similar to hexagonal ice \cite{finney2002structures, Mariedahl2018x}; however, at longer ranges the similarity drops. It remains unclear what structural model of LDA can best explain this.

Computer simulation can provide the atomic scale resolution needed to determine the structure of LDA.
A common way to obtain LDA in simulation is to quench liquid water rapidly to avoid crystallisation. 
Martelli et al. quenched the TIP4P/2005 water model with a cooling rate ($\kappa$) of 1 K ns$^{-1}$ to obtain LDA samples. They searched the LDA model for crystal-ice domains in the second coordination shell and found none, concluding it was indeed fully amorphous \cite{Martelli2018searching}. 
Furthermore, by quantifying the large-scale density fluctuations they showed it was nearly hyperuniform \cite{Martelli2017uniform}.
However -- as discussed by Limmer and Chandler, \cite{Limmer2014Theory} and others \cite{Giovambattista2004, Giovambattista2004b, Giovambattista2005, Gartner2021} -- $\kappa$ is a key parameter which when varied affects the structural and energetic properties of water glasses.
A range of different amorphous LDA samples is thus achievable in simulation.
Evidence of multiple configurations corresponding to LDA have also been observed via decompression or heating high-density amorphous ice (HDA) \cite{Giovambattista2003PRL, MartelliJPCB2023, Guillot2003Polyamorphism}.
Moreover, given the suggestions of a crystal-like structure in experiment, whether these fully amorphous states achievable in simulation are the best model for the LDA achieved in experiment (and detected in nature) is a key question to resolve.

\begin{figure*}
    \includegraphics[width=1.0\linewidth]{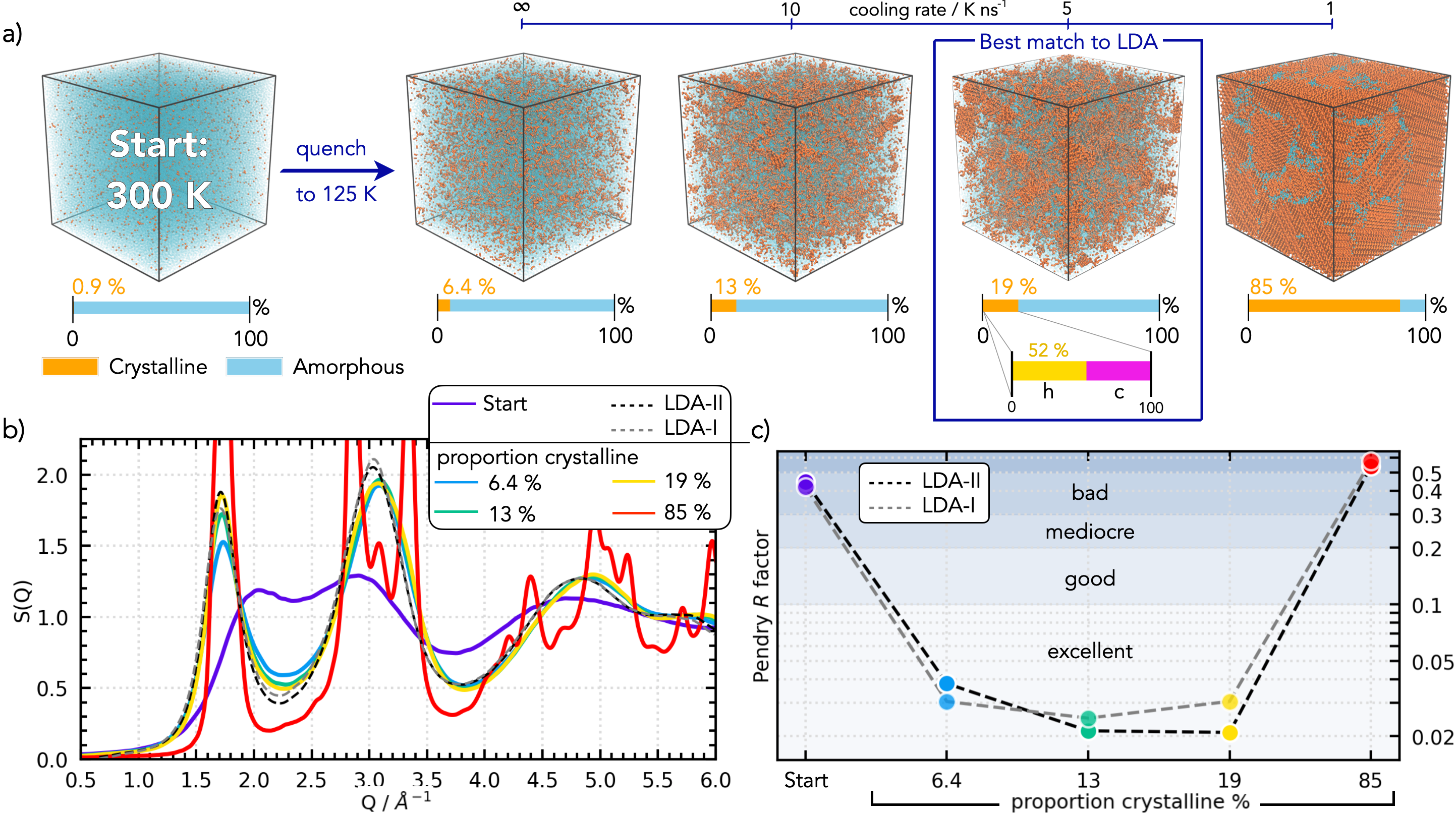}
    \caption{%
    \textbf{Quenching liquid water: the experimental diffraction data of LDA is best modelled with a partially crystalline state.}
    (a) Snapshots of systems taken from quenching water from 300 K to 125 K at different cooling rates. ``Start'' indicates equilibrated water at 300 K and $\infty$ K ns$^{-1}$ indicates immediate cooling to 125 K. 
    (b) $S(Q)$ of the quenched systems along with experimental data of LDA-II from ref. 44\textcolor{white}{\cite{Mariedahl2018x}} and LDA-I from ref. 54\textcolor{white}{\cite{Mariedahl2019RSC}}
    (c) Pendry R factor (of the agreement between experiment and theory) against crystallinity of structure achieved in quenching.
    The proportion of hexagonal (h) and cubic (c) local environments are indicated for the best model in (a).
    }
    \label{fig1}
\end{figure*}

In this work, we establish the structure of experimental LDA using computational and experimental approaches.
Agreement with the structure factor, $S(Q)$, of LDA -- as measured using wide-angle X-ray scattering of LDA-I and LDA-II by Mariedahl et al. (2018) and (2019) \cite{Mariedahl2018x, Mariedahl2019RSC} -- is screened across a range of computational models obtained from two diametrically opposed pathways: (i) quenching liquid water at different $\kappa$; and (ii) constructing large-scale polycrystalline ice structures using Voronoi domains followed by structural relaxation.
In the former, the structure of LDA is explored starting from the disordered liquid water state, whilst in the latter it is approached using ordered crystalline ice building blocks. 
For optimal accuracy, we consider stacking disorder in the building blocks.
And computationally large system sizes (on the order of 190,000 -- 320,000 water molecules) are employed to better enable capturing the short-range order vs. long-range anisotropy in polycrystalline systems.
Experimentally, we investigate the crystallisation products of different types of LDA with respect to stacking disorder. The aim is to search for memory effects: these are expected if LDA is partially crystalline but not if it is fully amorphous.

\section{Results}

\subsection{Quenching liquid water indicates that LDA-I and LDA-II are partially crystalline states.}

To begin, we performed molecular dynamics simulations in which liquid water, represented by the mW water model \cite{mWmodel}, was quenched from 300 K to 125 K (see \textit{Methods}). 
Depending on the cooling rate ($\kappa$) a wide variety of amorphous ice materials was obtained, ranging from partially crystallised states to polycrystalline (almost fully crystallised) states, as shown in Fig. \ref{fig1} a.
The former are observed for $\kappa$ $\geq$ 5 K ns$^{-1}$, and resemble multiple isolated grains of crystalline ice embedded within an amorphous matrix.
The latter is obtained with $\kappa =$ 1 K ns$^{-1}$, and resembles multiple grains of ice packed together.
The formation of these various structures simply comes down to a competition of timescales between water diffusion and ice crystallisation. 
The key question is which, if any, of these amorphous and partially crystalline states is a good model for LDA?

To help answer this, we report in Fig. \ref{fig1}b the structure factors of the quenched systems alongside experimental measurements of LDA.
In experiment, Winkel et al. reported the existence of two sub-states of LDA -- which they named LDA-I and LDA-II -- that exhibit subtle structural differences in their diffraction patterns \cite{Winkel2009LDA2}.
It has been suggested that LDA-I may contain crystalline ice whilst LDA-II does not on the basis of crystallisation temperatures \cite{Elsaesser2010, Seidl2013}.
We therefore compare the X-ray diffraction pattern of LDA-II collected by Mariedahl et al. (2018) \cite{Mariedahl2018x} and the X-ray diffraction patterns of LDA-I by Mariedahl et al. (2019) \cite{Mariedahl2019RSC} to our computational models.

Fig. \ref{fig1}b shows that the optimal structural model for both LDA-I and LDA-II is a partially crystalline state -- yet with stacking-disordered ice -- present in numerous and diversely sized grains.
Interestingly, agreement with the experimental data is observed to be a ``goldilocks'' scenario with the best model being not too amorphous and not too crystalline.
The key effect of crystallinity in these states is on the prominence of the first peak and trough in $S(Q)$, which reflects more long-range structural characteristics compared to the higher \textit{Q} features.
The nucleation rate has a per unit time and per unit volume dependence, and these dependencies can physically explain the states achieved from quenching:
the former explains the effect of $\kappa$;
the latter is captured by system size. 
Fig. S1 shows that smaller systems give poorer matches to LDA as a result of the reduced length scale impairing the ability to capture polycrystalline structures.
The proportion of cubic (\textit{c}) local environments in stacking disordered ice I -- termed the ``cubicity'' -- in the ice grains is 48 \%, which is close to the expected 50 \% cubicity obtained in homogeneous ice nucleation \cite{Malkin2012, Moore2011, AkbariPNAS2015, Lupi2017, DaviesPNAS2021}.

To quantify the agreement between the computational models and experiment the Pendry R-factor ($R$) was employed \cite{Pendry1980} over the experimental $Q$ range of $1.0-23$ $\text{\AA}^{-1}$.
$R$ was developed and is widely used in low-energy electron diffraction to quantify the agreement of computational models with experimental data (see \textit{Methods}). A value of $R$ = 0 represents perfect agreement whereas values of $\approx$ 0.1 are taken as excellent, $\approx$ 0.2 as good, $\approx$ 0.3 as mediocre, and $\approx$ 0.5 as bad \cite{Zanazzi1977}.
Fig. \ref{fig1}c shows that the three partially crystallised states have excellent agreement with experiment.
The best models for LDA-I and LDA-II are the systems with 13 \% and 19 \% crystalline ice, respectively.
The respective $R$ values are 0.025 and 0.021, which indicate extremely high agreement.
Fig. S2 shows qualitative agreement between $R$ and the coefficient of determination.
Fig. S3 shows $S(Q)$ for the best LDA model and experimental data over the full $0.65-23$ $\text{\AA}^{-1}$\ range.

Our results suggest that LDA-I and LDA-II are both partially crystalline states, containing stacking disordered ice.
Furthermore, we find that the optimal match for LDA-I is slightly less crystalline than that found for LDA-II -- in contrast to the suggestion in literature that LDA-I contains crystallites whereas LDA-II does not (on the basis of crystallisation temperatures).
To investigate this further, Fig. S3 shows an LDA-I diffraction pattern along with two different experimental measurements of LDA-II.
The LDA-I data lies within the variation of the two different LDA-II measurements. This suggests the difference between LDA-I and LDA-II may be within experimental error.
Furthermore, in Raman spectroscopy no difference could be found between LDA-I and LDA-II \cite{Shephard2013anneal}.
In conclusion, we find LDA-I and LDA-II are both similar partially crystalline states, with any structural differences between LDA-I and LDA-II due to subtle differences in crystallinity, experimental error or both.
Given this, the rest of this study employs the LDA-II experimental sample of ref. 44\textcolor{white}{\cite{Mariedahl2018x}} as ``LDA'' for the sake of brevity.

\begin{figure}[]
    \centering
    \includegraphics[width=1.0\linewidth]{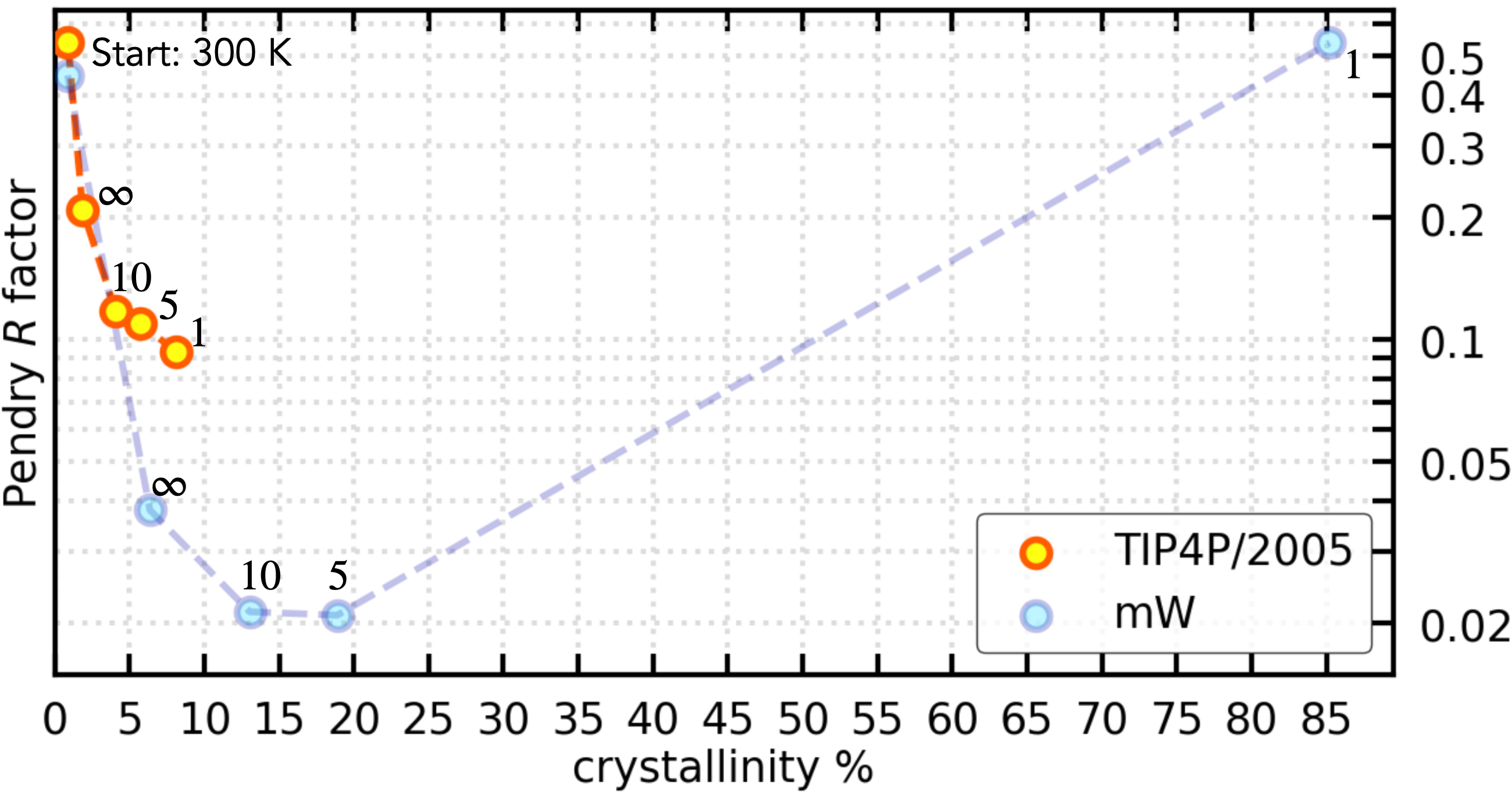}
    \caption{
    \textbf{
    Quenching two water models indicates LDA is a partially crystalline state.} Pendry $R$ factor of experimental LDA against percentage crystallinity of computational model, for the TIP4P/2005 and mW water models, as obtained by the quenching procedure detailed in Fig. \ref{fig1}. Data points are labelled with the employed cooling rate ($\kappa$). 
    }
    \label{fig2}
\end{figure}

\subsection{LDA is optimally modelled as a partially crystalline state by both the mW and TIP4P/2005 water models.}

We next extended the analysis by applying the same quenching protocol to the the atomistic TIP4P/2005 water model \cite{TIP4P2005}.
In Fig. \ref{fig2}, we show that TIP4P/2005 and mW optimally model experimental LDA as a partially crystalline state.
TIP4P/2005 follows the same pattern as mW, displaying an improved match to LDA as the proportion of crystallinity increases from 1.9 \% to 4.1 \% to 5.7 \% to 8.1 \%.
Fig. S4 show the respective system snapshots and $S(Q)$.

The slower nucleation rate and dynamics of TIP4P/2005 result in more highly amorphous states, when subject to the same cooling ramp procedure as mW.
The highest crystallinity achieved by TIP4P/2005 in the cooling ramps yielded the best match to LDA, thus 8.1 \% is a lower bound estimate for LDA's crystallinity for TIP4P/2005. 
The trend indicates that higher levels of crystallinity would yield even more accurate agreement, however the computational cost of obtaining this is not practical in this study due to the model's order of magnitude increase in cost combined with the $\kappa$ required (on the order of 0.01 K ns$^{-1}$).
We refer the reader to the methods section for more discussion on the water models.


\begin{figure*}
    \includegraphics[width=1.0\linewidth]{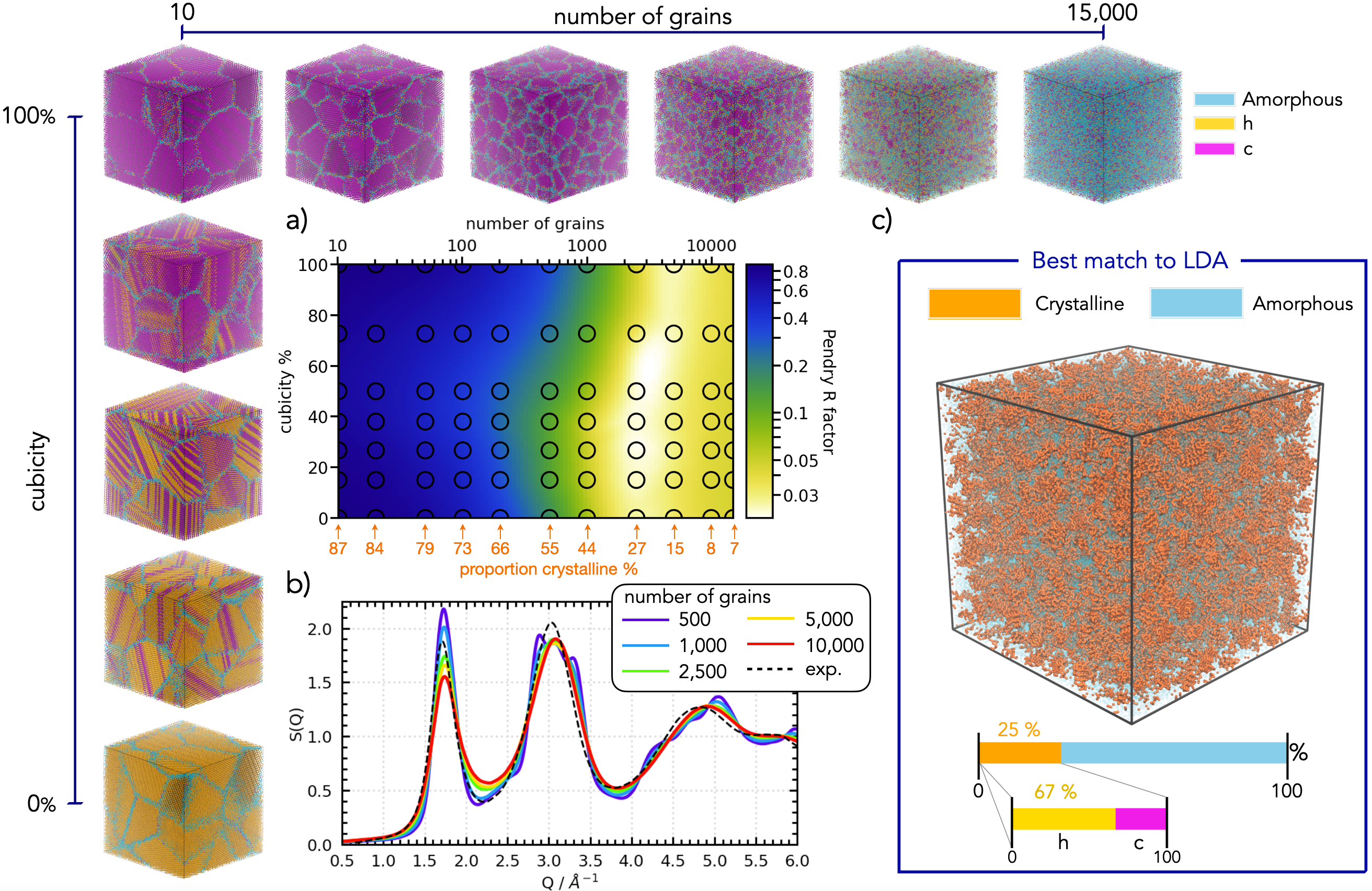}
    \caption{\textbf{Constructing LDA with Voronoi domains containing crystalline ice.}
    The agreement between experiment and polycrystalline systems was explored for model systems across a grid of different levels of cubicity (of the subset of molecules in crystalline environments) and polycrystallinity (given by the number of grains). Snapshots of the grid's top and left boundaries decorate the borders of the figure, and the approximate mean diameters of the grains are shown in Fig. S5.
    (a) Heatmap of the Pendry $R$ factor for the agreement of the $S(Q)$ between experiment and the model systems and the computational systems as a function of the proportion of molecules with crystalline ice environments and the cubicity.
    (b) The computational $S(Q)$ with optimal agreement with experiment for different levels of polycrystallinity. 
    (c) Snapshot of the system with the best match to the experimental data, with its local atomic environments indicated. System contains approximately 320,000 water molecules.
    }
    \label{fig3}
\end{figure*}

\subsection{Models of polycrystalline ice indicate LDA is a partially crystalline state.}

Next, we investigated the diametrically opposed pathway to creating LDA from liquid water; specifically, creating LDA from crystalline ice. 
The random arrangement of many ice grains can broaden the diffraction features of crystalline samples, yielding amorphous-like $S(Q)$.
To this end, large-scale ($\approx$ 320,000 water molecules) periodic polycrystalline structures with different levels of polycrystallinity and cubicity were built using Voronoi domains.
The structures were subsequently geometry optimised and annealed at 125 K using the mW water model.
A grid search was set up as illustrated in Fig. \ref{fig3} to find the most optimal model for LDA. 

Fig. \ref{fig3}a shows a heatmap of the $R$ values of the relaxed polycrystalline models benchmarked against the experimental LDA $S(Q)$.
A strong dependence of the agreement is seen with polycrystallinity.
Fig. \ref{fig3}b shows this is due to the increasing amorphous nature of the diffraction patterns with increasing polycrystallinity. 
The ``goldilocks'' scenario is again observed.
The optimal match is found for the 2,500 grain systems which have an optimal $R$ value of 0.022.
Both lower and higher levels of polycrystallinity (and thus proportion of crystalline versus amorphous) result in poorer matches to experiment.
Fig. \ref{fig3}c shows the structure of the best fit. Again, the structure is a partially crystallised state with isolated grains of crystalline (yet stacking disordered) ice surrounded by amorphous ``filler'' regions.
The amorphous regions arise during the structure relaxation due to the lattice mismatches between the various Voronoi domains. 
Since ice is a ``soft'' molecular material it yields quite thick amorphous regions at the grain boundaries.
At this high level of polycrystallinity the system is 25 \% crystalline and 75 \% amorphous.
It thus no longer resembles a ``traditional'' polycrystalline system -- the type highlighted by Amann-Winkel et al. to be of interest but not a probable structure for amorphous ices \cite{Amann2016} -- which have sharp interfaces at the grain boundaries and are therefore primarily crystalline.
A weak, but noticeable, dependency on the match to experiment is seen for the level of cubicity in Fig. \ref{fig3}a, with a best match around $25 - 50$ \%.

\begin{figure*}[!htp]
    \includegraphics[width=1.0\linewidth]{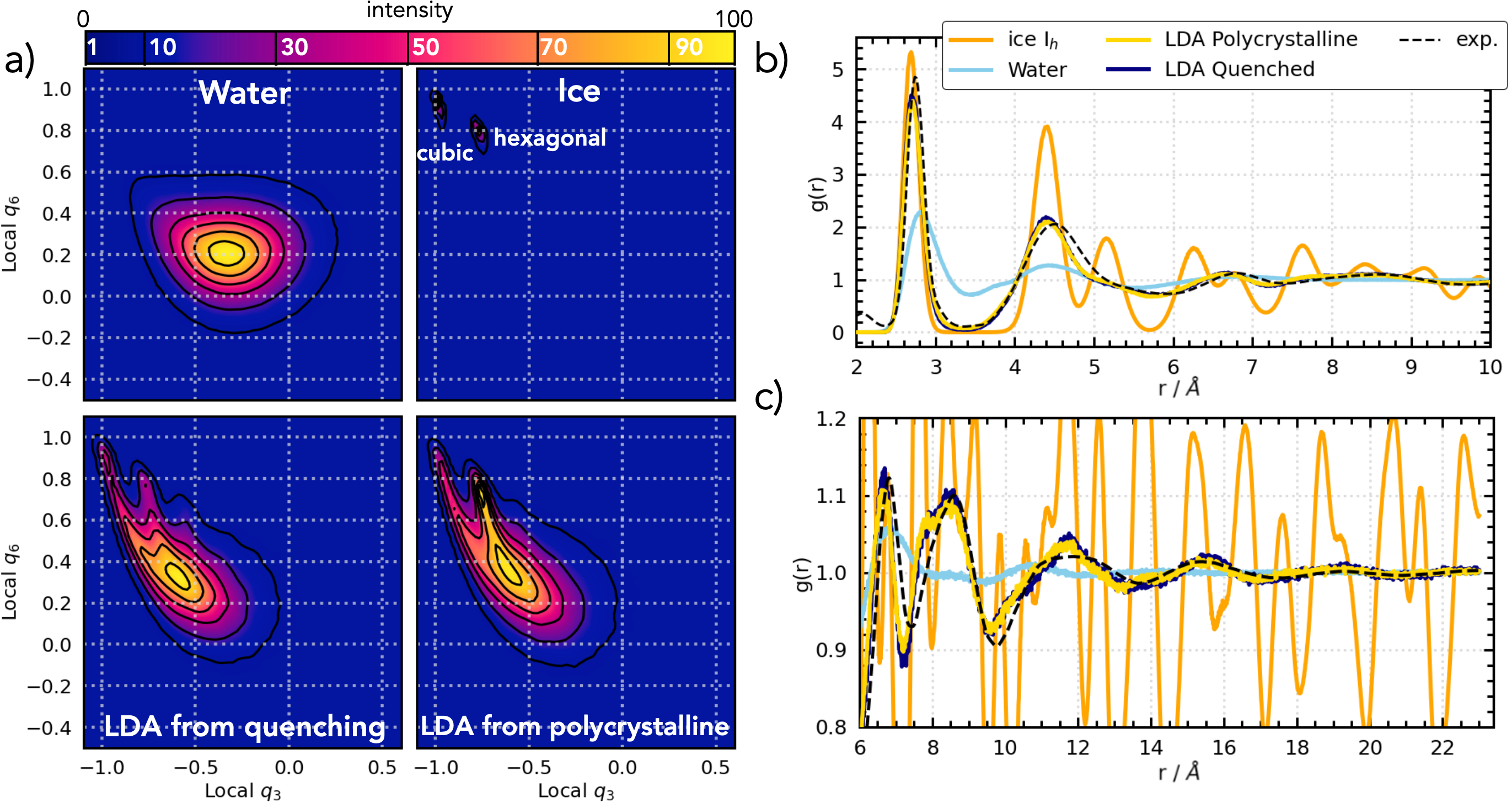}
    \caption{%
    \textbf{The hybrid amorphous-crystalline nature of LDA.}
    (a) Probability densities of the local atomic environment of water, ice, and the LDA models. Contours are drawn at 90, 70, 50, 30, 10, and 1 \% of the height of the respective probability densities.
    (b) \& (c) Oxygen-oxygen radial distribution functions at short (b) and long (c) range for the computational models and experiment.
    Ice I$_{\text{h}}$, ice I$_{\text{c}}$ and the LDA models are sampled at 125 K and water at 250 K. 
    }
    \label{fig4}
\end{figure*}

The two different approaches -- building polycrystalline ice and quenching liquid water -- converge to very similar structural models of LDA.
This gives additional confidence that experimental LDA is indeed partially crystalline.
The subtle differences between the two models of LDA are discussed in the SI.
Here, it is sufficient to say that differences in the uniformity of grain size are the origin of the slight difference in the crystallinity.

\subsection{Exploring the hybrid amorphous-crystalline nature of LDA.}

We next further investigated the hybrid amorphous-crystalline nature of the optimal LDA models from quenching liquid water and building polycrystalline ice.
In Fig. \ref{fig4}a the distributions of the atomic local environments of water, ice I$_{\text{h}}$, ice I$_{\text{c}}$, and the LDA models are plotted.
``Local'' here refers to measurements of order within the first coordination shell, as defined by local $q_{6}$ and local $q_{3}$ (see \textit{Methods}).
The local environments of the two LDA models are very similar to each other.
They contain significant amounts of highly crystalline and amorphous (as explored by liquid water) atomic local environments, with a smooth distribution of environments explored between these two levels of local order. 

The hybrid amorphous-crystalline nature is also reflected in the oxygen-oxygen radial distribution functions, $g(r)$, shown in Fig. \ref{fig4}b and c.
The two LDA models have near identical $g(r)$ and both show a very good agreement to experiment, which is well maintained up to the longest range signal measured in experiment of 23 \AA.
In the short-range, up to $\sim 6$ \AA\, the first coordination shell of LDA is very similar to crystalline ice whereas the second coordination shell is broadened.
From $\sim 6$ \AA\ onwards LDA's radial correlations quickly decay leaving it dissimilar to crystalline ice; yet, the correlations remain much stronger than in liquid water. Thus, again LDA lies between the two.
Mariedahl et al. (2018) made similar observations from their experimental signal, stating ``it is essential to develop models of amorphous ices that can better describe these intermediate-range correlations’’ \cite{Mariedahl2018x}.
Similar observations were also made by Finney et al. \cite{finney2002structures} who found ``considerable similarity in the local order between LDA and ice I\textsubscript{h}..., [but] beyond the second shell there are increasing discrepancies between their respective $g(r)$''.
The strong agreement between the LDA models and experiment show this structure is accurately modelled and explained by a partially crystalline state.
The anisotropically arranged grains result in short-range similarity to crystalline ice, but long-range decays in $g(r)$.
The primitive ring distributions of the LDA models are shown to be very similar in Fig. S6, and are also consistent with partially crystallised states.

\begin{figure}[!h]
    \centering
    \includegraphics[width=1.0\linewidth]{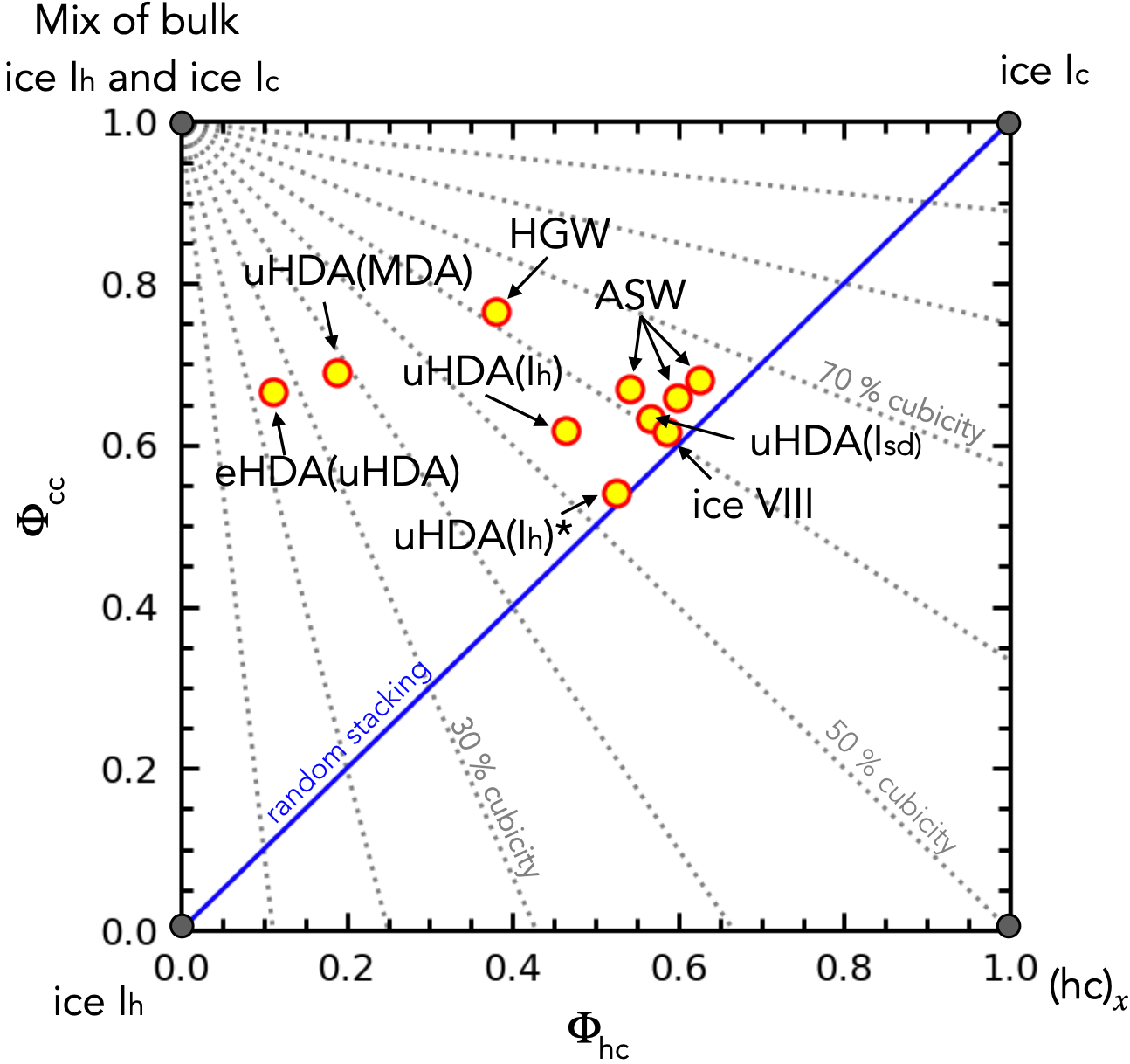}
    \caption{
    \textbf{Experimental stacking probabilities of ice I$_{\text{sd}}$ obtained from LDA depend on how the LDA was made.}
    In this ``stackogram'', $\Phi_{\text{cc}}$ and $\Phi_\text{{hc}}$ define the probabilities of cubic stacking after a previous cubic or hexagonal stacking event, respectively.
    The blue diagonal line represents random stacking, where $\Phi_{\text{hc}}$ and $\Phi_{\text{cc}}$ are equal.
    The dashed grey lines indicate constant cubicities.
    The four corners represent ice I$_{\text{h}}$, ice I$_{\text{c}}$, a physical mixture of the two and the (hc)$_{x}$ polytype which consists of strictly alternating cubic and hexagonal layers of ice.
    For each data point the ``parent'' phase from which the LDA was made is noted. 
    For uHDA and eHDA the ``grandparent'' phases are also given in brackets. uHDA(I$_{\text{h}})\ast$ was created using a different experimental procedure to the other uHDA samples (see \textit{Methods}). 
    }
    \label{fig5}
\end{figure}

\subsection{Cubicity memory effects in experimental LDA samples can be explained by a hybrid amorphous-crystalline state.}

Experimentally, we found indirect evidence for the presence of crystallinity in LDA by heating it to form stacking disordered ice (ice I$_{\text{sd}}$) and quantifying the stacking probabilities.
Fig. \ref{fig5} shows a remarkable variability in the stacking probabilities of ice I$_{\text{sd}}$ obtained from differently prepared LDA samples.
Each LDA sample has a different ``parent'' phase: unannealed high-density amorphous ice (uHDA) formed by compressing ice I$_{\text{h}}$, ice I$_{\text{sd}}$ and medium density amorphous ice (MDA) respectively \cite{MDA}; expanded high-density amorphous ice (eHDA) formed by heating a compressed sample of uHDA; ice VIII; vapour deposition to give amorphous solid water (ASW); and the hyperquenching, performed by Kohl et al. in ref. 67,\textcolor{white}{\cite{Kohl2000hgw}} of an aerosol of liquid water droplets to give hyperquenched glassy water (HGW).

A fully amorphous LDA structure should not display memory effects with respect to the parent phase.
Instead, nucleation occurring from a fully amorphous structure would follow homogeneous nucleation, which results in random stacking with 50 \% cubicity \cite{Malkin2012, Moore2011, AkbariPNAS2015, Lupi2017, DaviesPNAS2021}.
However, the obtained cubicities here range from $20-65$ \%.
This indicates that that crystal growth from already existing ice I\textsubscript{sd} seeds is the dominant process, not homogeneous nucleation.
In summary, if LDA is partially crystalline, then the route to formation can be expected to potentially result in different nanocrystallite cubicities, and thus give rise to different cubicities upon recrystallisation. This is observed.
A fully amorphous material would not display such memory effects.

The relatively weak dependence of cubicity -- observed earlier in the polycrystalline LDA models -- shows that crystalline ice I grains with different cubicities can be well hidden in the broad diffraction features of LDA samples.
Similar differences in the crystallisation behaviour of LDA samples created via different pathways have also been reported by Seidl et al. \cite{Seidl2013} and Mariedahl et al. \cite{Mariedahl2019RSC} -- but accurate cubicities have not been reported.

\section{Discussion}

In this work, we present a combined computational and experimental approach showing that LDA has a hybrid amorphous-crystalline structure.
Computational models with near exact agreement of the structure factors with experiment were achieved through multiple modelling approaches.
The quenching of liquid water -- both with the mW and TIP4P/2005 water models -- optimally modelled LDA as a partially crystalline state.
And building polycrystalline ice systems -- a diametrically opposed pathway to quenching -- also optimally modelled LDA as a partially crystalline state.
In all cases, matching experiment was found to be a ``goldilocks'' scenario, where the structure is not too amorphous and not too crystalline.
The models resemble a partially crystalline state with grains of stacking disordered crystalline ice surrounded by amorphous ``filler regions''.
Experimentally, indirect evidence of LDA's hybrid amorphous-crystalline state was found by transforming LDA to ice I$_{\text{sd}}$: a memory effect with respect to the parent material of LDA is seen in the stacking probabilities that cannot be explained by a fully amorphous LDA state and hence indicates the presence of crystalline ice grains already in the LDA with different cubicities.

Finney et al. estimated that any crystallites in LDA were less than 15 \AA\ \cite{finney2002structures}, and used this to suggest ruling out a partially crystalline structure of LDA.
Instead, we find that a partially crystalline structure is indeed the optimal model for LDA, as it provides the required short-scale local crystalline structure but long-scale homogeneity.
The distribution of many (anisotropically arranged) small crystallites has been discovered to be essential to modelling LDA. 
The resulting structure is far from a fully amorphous material: both at the local- (individual crystallites) and global-scale ($\approx$ 25 \% crystalline).
The system snapshots in Fig. \ref{fig1} and \ref{fig3} perhaps best illustrate this partially crystalline nature.
An estimate for the diameter of the grains in these systems is provided in Fig. S5.
We find the optimal model system, containing 2,500 grains, has a mean grain diameter of 25 \AA. 
The crystalline sizes are subject to stochastic variation. Therefore, using the grain sizes from the next nearest models to the optimal, we suggest crystallites in LDA will range from $30 - 15$ \AA\ in size.
It is worth noting, Finney et al. considered only ice I\textsubscript{h} crystallites and did not consider for stacking disorder, as done here, which has a broadening effect on diffraction features.

The mW water model enabled the discovery of the optimal models, as its computational cost and dynamics enabled the system sizes and time scales required to capture the partially crystalline systems.
mW has been widely and successfully employed to study water, crystalline ice and amorphous ice. 
And has been specifically shown to accurately model low-pressure states of ice, including LDA \cite{mWmodel}.
However, it is a coarse-grained model meaning it has inherent limitations such as the ability to distinguish between acceptors and donors of hydrogen bonds.
We therefore also employed the atomistic TIP4P/2005 water model, a well regarded classical water model that has been extensively used to model amorphous ice.
TIP4P/2005 also optimally modelled LDA as partially crystalline. The increased cost meant more highly crystalline structures were not feasible to obtain; however, it was shown to be at least 8\% stacking disordered crystalline ice, with a clear trend to indicate the optimal models would have higher crystallinity.
Two widely used and disparate water models have thus been shown in this study to optimally model experimental LDA as a partially crystalline state.
Looking to the future, extending the analysis to machine learned force-fields for water is desirable. However, the time scales and system sizes necessary are far beyond current feasible computational costs at present.

Different preparation routes yield different nomenclatures for LDA in the literature.
Of particular importance to this study is LDA-I and LDA-II, since they had been suggested to be two distinct sub-states of LDA \cite{Elsaesser2010, Seidl2013}.
We found that both LDA-I and LDA-II are accurately modelled as similar partially crystalline states. Any difference between the two likely comes from experimental error, and/or potentially subtle differences in crystallinity and/or cubicity.
Moreover, if LDA is a partially crystalline material then different preparation routes should lead to potentially different cubicities in the nanocrystallites, and thus different behaviours upon crystallisation to ice I. This is observed experimentally (over seven different formation routes to LDA) and thus provides evidence that the many different nomenclatures of LDA in the literature are indeed partially crystalline materials.

In addition to experimental X-ray diffraction results, the very similar local environments in LDA and ice I presented here are also consistent with their Raman spectra of their O-H stretching modes \cite{Carr2014}. In fact, the LDA to ice I$_{\text{sd}}$ phase transition is barely visible in Raman spectroscopy.
Moreover, the very small downshift observed in the O-H stretching modes upon crystallisation is consistent with strain release. Such strain release would occur between the anisotropically strained ice grains in LDA due to the lattice mismatch between neighbouring grains.
The strain is released when LDA crystallises to ice I\textsubscript{sd} as the ice grains grow larger, coalesce and align in orientation.
In Fig. S7, the vibrational density of states for the different levels of polycrystallinity explored in this study are plotted. This shows a qualitative agreement with experiment whereby the spectra show a weak dependence on the level of polycrystallinity and no dependence on the cubicity.

A hybrid-crystalline structure of LDA has potential wide-reaching implications.
Being the most abundant form of water in the universe, LDA is important in astronomy; for instance, in astrochemistry LDA is thought to play a key role in promoting the variety, complexity and richness of molecules observed in the stellar-forming regions of space \cite{Burke2010, Ehrenfreund2000}.
In cryopreservation, the vitrification (avoidance of ice crystallisation) of biological matter is thought to be key to cell survival -- here a fully amorphous LDA is desired to act as the embedding matrix \cite{Huebinger2016, NatCryoReview}.
A fully amorphous LDA is similarly desired in the widely used technique of cryogenic electron microscopy \cite{Cressey2017} -- achieving a truly amorphous LDA would help prevent sample damage and improve resolution. 
Crucially, LDA is key for our understanding of liquid water. A corresponding liquid state to LDA is central to the two-state model. And whilst a fully amorphous LDA can be achieved in simulation, confirming if a fully amorphous LDA can be achieved experimentally or exists in nature is now of paramount importance -- especially given the recent discovery of MDA \cite{MDA}. 
To achieve a fully amorphous LDA, novel techniques might need to be employed. For instance, a rapid enough cooling rate might completely avoid crystallisation.

Finally, the findings of this study have implications for the structural nature of other amorphous materials. 
The other amorphous ices, such as HDA, are of immediate interest.
As are the myriad of glasses that underpin many technologies (for example, OLEDs and fibre optics). The beneficial application of glasses over crystals is largely due to their macroscopic homogeneity and absence of grain boundaries \cite{Ediger2017perspective}. 
Determining whether materials are truly glassy or not could provide future opportunities for technological innovation.
Indeed, in the literature it has already been shown that amorphous silicon can be equally well modelled by a partially crystalline structure instead of one that is fully amorphous \cite{Rudee1972silicon, Treacy2012silicon}.
Another question that arises is the nature of ``ultrastable glasses'' \cite{Swallen2007organic, Ediger2017perspective} -- a class of material that can exhibit highly desirable properties. 
The low entropy of these materials could be explained by the presence of crystalline grains that were previously undetected, thereby providing a resolution to the Kauzmann paradox \cite{Kauzmann1948nature}. 

\section{Methods}

\subsection{Molecular dynamics (MD) simulations.}
MD simulations were performed with the large-scale atomic/molecular massively parallel simulator (LAMMPS) code \cite{Plimpton1995}. Cubic simulations boxes periodic in $x,y,z$ were used. The constant number of molecules, constant pressure, and constant temperature (NPT) canonical ensemble was sampled using 10-fold Nos\'e-Hoover thermostats and barostats.
MD simulations of mW used thermostats and barostats with relaxation times of 500 femtoseconds (fs) and 5 picoseconds (ps) respectively, and integrated the equations of motion with a timestep of 10 fs.
MD simulations of TIP4P/2005 used thermostats and barostats with relaxation times of 200 fs and 2 ps, respectively.
Simulations of quenching liquid water and of annealing polycrystalline ice employed isotropic and anisotropic barostats, respectively.

There are, of course, many water models. 
In this study we chose to use mW because it offers the best trade-off between accuracy and computational cost. The choice was based on several reasons:
(i) mW accurately captures important properties of water such as the density, structure, melting temperature, and the formation of LDA \cite{mWmodel};
(ii) it has been widely and successfully used in studying the nucleation of ice \cite{mWmodel, Li2011, Li2013, Lupi2014, Lupi2017, Sosso2016b, ManyFaces, DaviesPNAS2021, DaviesPNAS2022};
(iii) it accurately captures the metastability of ice I$_{\text{c}}$ \cite{Quigley2014,mWmodel} as well as the thermodynamics of stacking faults \cite{Lupi2017, Hudait2016,Hondoh1983};
(iv) crucially, because it is considerably cheaper than an all-atom model and has faster dynamics mW enables quenching and crystallisation simulations in large system sizes ($> 100,000$ particles) -- such system sizes are essential for determining the balance of ice-like and amorphous domains. 
The TIP4P/2005 water model was employed as an atomistic water model as it has also been widely and successfully applied to study water and ice, including the formation of LDA \cite{TIP4P2005, TIP4PIceTmd, Sosso2016b, Martelli2017uniform, Martelli2018searching, Debenedetti2020Science}. 

Quenching simulations were performed in a similar fashion to Gartner et al. \cite{Gartner2021}. Water was first equilibrated at 300 K and 0 Pa for 2 ns, prior to generating starting configurations.
The desired $\kappa$ rate was achieved by smoothly ramping the temperature from 300 K to 125 K over the required time period. For $\kappa = \infty$ K/ns, the structure was further relaxed (4 ns for mW, and 20 ns for TIP4P/2005).

Using our Stacky program \cite{Stacky}, a range of ice I supercell structures with 60 layers, 322,560 mW water molecules and cubicities of 1, 0.73, 0.5, 0.37, 0.27, 0.13 and 0 were produced. After transforming the hexagonal to orthorhombic cells, the cell dimensions were 21.5, 21.8 and 22.0 nm in \textit{x}, \textit{y}, \textit{z}.
The purpose-written nanoVD program then used these structures to create 5, 10, 20, 50, 100, 200, 500, 1,000, 2,500, 5,000, 10,000 or 15,000 Voronoi domains within a cell of the same size.
The center points of the Voronoi domains were randomly chosen. 
The ice I structures were rotated randomly using a \textit{zyz} rotation matrix and shifted by random fractions of the cell edges. The latter step is essential to ensure that the final polycrystalline structure reflects the cubicity of the parent structure. The structures were then placed at the centers of the Voronoi domains followed by deleting all molecules that were more than half-way away with respect to the centers of the surrounding Voronoi domains including those located outside the cell through the periodic boundary conditions. 
As a quality check, it was confirmed for each polycrystalline structure that the number of the mW molecules approximately matched the corresponding number in the start structure.

The various polycrystalline structures with different domain sizes and cubicities were then geometry optimised in LAMMPS using the steepest descent algorithm, then equilibrated at 125 K and 0 Pa for 50 ps with a timestep of 0.25 fs, then for 2 ns with a timestep of 10 fs, prior to undertaking analysis.
Fig. S10 shows snapshots of the systems before and after the geometry optimisation plus equilibration.

Structure factors were calculated using the Debyer software (https://github.com/wojdyr/debyer) with a cutoff of half the minimum box dimension, and the sinc-fuction (sin(x)/x) to dampen the cut-off and reduce the Fourier termination effects.
Ring statistics were calculated with the Rigorous Investigation of Networks Generated using Simulations (R.I.N.G.S) software \cite{RINGS} and the primitive rings criterion.
The Visual Molecular Dynamics software \cite{Humphrey1996} was used to calculate radial distribution functions and to generate images of systems \cite{STON1998}.

\subsection{Classifying local environment of water molecules.}
The local $q_{6}$ order parameter of Li et al. was used to classify mW molecules as either crystalline or amorphous \cite{Li2011}.
The threshold for labelling a particle as crystalline or amorphous was calculated following the methodology used by Espinosa et al. \cite{VegaSeedingApproach}.
Periodic boxes of bulk ice I$_{\text{h}}$, ice I$_{\text{c}}$, and water containing 184,000, 216,000 and 192,000 mW molecules respectively were simulated at 250 K and 0 Pa, and the corresponding local $q_{6}$ values were extracted every 200 ps over 2 ns. 
The distribution of the local $q_{6}$ values for crystalline and amorphous environments were then taken from the bulk crystalline ice and bulk liquid water simulations respectively.
As shown in Fig. S11 the cutoff for the local $q_{6}$ between crystalline and amorphous environments are then taken where the probability of mislabelling a particle as crystalline when it is amorphous and vice versa is equal: local $q_{6}$ gives a mislabelling percentage of 0.8 \% here.
A similar methodology was used to classify molecules as either ice I$_{\text{h}}$ or ice I$_{\text{c}}$, where instead the local $q_{3}$ values were compared between bulk ice I$_{\text{h}}$ and ice I$_{\text{c}}$. 
Fig. S12 which shows that local $q_{3}$ gives a mislabelling percentage of 0.04 \% here for the mW model.

The same method was used to distinguish crystalline form amorphous molecules for the TIP4P/2005 model (using the configuration of the oxygens); however an alternative definition of local $q_{6}$ by Lechner and Dellago \cite{Lechner2008} was used as it was found to give a smaller mislabelling percentage here than that of Li et al. Fig. S13 shows a mislabelling percentage of 0.84 \% was achieved.

The plugin for metadynamics 2 (PLUMED2) \cite{Plumed_1,Plumed_2} was used to calculate local $q_{6}$ and local $q_{3}$.

\subsection{Pendry R-factor.}
The Pendry reliability factor (``Pendry R-factor'') was calculated as defined by Pendry \cite{Pendry1980}.
Let $f$ be a signal and $\dot{f}$ be its first derivative.
To compare $f$ with another signal, the logarithmic derivative is employed to give emphasis on the location of the maxima/minima rather than the amplitudes.
\begin{equation} \label{L}
\begin{split}
L & = \dot{f}/f 
\end{split}
\end{equation}
A comparison between theory, $\bar{L}$, and experiment, $\hat{L}$, using the L functions gives too high an emphasis to zeros; therefore, the Pendry y-function is then employed to give a similar emphasis to troughs and peaks in the signal
\begin{equation} \label{Y}
\begin{split}
Y & = L^{-1}/(L^{-2} + V^{2})
\end{split}
\end{equation}
The imaginary part of the electron self energy term used in the LEED community, V, was set to two here.
The Pendry R factor between theory, $\bar{Y}$, and experiment, $\hat{Y}$, is then defined as
\begin{equation} \label{R}
\begin{split}
R & = \frac{\Sigma_{i}(\bar{Y}_{i} - \hat{Y}_{i})}{\Sigma_{i}(\bar{Y}_{i}^{2} + \hat{Y}_{i}^{2})} 
\end{split}
\end{equation}
This was employed between computational and experimental $S(Q)$ over the experimental Q range of $1.0-23$ $\text{\AA}^{-1}$.
$R$ was employed due to its popularity in the LEED community in quantifying the agreement of computational models with experiment. 
In Fig. S1, the results obtained with $R$ are shown to qualitatively agree with the coefficient of determination in this study.

\subsection{Experimental procedures for creating the LDA ``parent'' phases.}
All uHDA samples, except from uHDA(I$_{\text{h}}$)*, were prepared in a 0.8 cm diameter pressure die lined with indium, pre-cooled to 77 K, and compressed to 1.6 GPa at 5 kN/min.
uHDA(I$_{\text{h}}$)* was prepared by injecting 0.4 mL H\textsubscript{2}O  into the pre-cooled indium cup, forming ice I$_{\text{h}}$, followed by pressure-induced amorphization (PIA) similar to Mishima et al. \cite{Mishima1984Nature}.
In the case of uHDA from MDA, a 77 K-cooled indium cup was filled with MDA and placed into the pressure die where the preparation of MDA can be found in ref. 66.\textcolor{white}{\cite{MDA}}
To create the parent phases of ice  I$_{\text{h}}$ and ice I$_{\text{sd}}$, the 0.4 mL H2O (Milli-Q, Millipore) injected into the cold pressure die was compressed to 0.35 GPa and heated to 200 K forming ice II. 
Following quenching to 77 K, ice II was decompressed to 100 N whereafter ice I$_{\text{sd}}$ or I$_{\text{h}}$ was extracted at 182 K or 260 K, respectively as determined by the in situ volume changes during heating. 
The resultant ice was then transformed to uHDA through PIA.
Further steps were taken when creating eHDA, which was made by heating uHDA to 145 K at 0.25 GPa before quenching to 77 K.
All samples were extracted in a liquid nitrogen environment for further analysis.

\subsection{Experimental procedure for transforming LDA samples to ice I$_{\text{sd}}$ upon heating.}
The various samples were transferred under liquid nitrogen into a purpose-built Kapton window sample holder mounted on a Stoe Stadi P diffractometer with Cu K$\alpha$1 radiation at 40 kV, 30 mA and monochromated by a Ge 111 crystal.
A Mythen 1K area detector collected data every 10 K from $93 - 260$ K with a heating rate of 1 K/min as controlled by an Oxford Instruments Cryojet HT.
The samples form LDA around 110 K (except ASW and HGW which are already LDA to begin with), and ice I$_{\text{sd}}$ around $140 - 150$ K, before finally converting to ice I$_{\text{h}}$ above 220 K.
An example of this heating procedure when applied to uHDA is shown in Fig. S8.
The first ice I$_{\text{sd}}$ patterns after the conversion from LDA upon heating were fitted with the MCDIFFaX software \cite{Salzmann2015mcdiffax}.
The software searches for the optimum lattice constants, stacking probabilities, peak profile parameters and zero-shift to reach a $\chi^{2}$ convergence and hence best-fit to the experimental data. The resultant fits overlaying the experimental data are shown in Fig. S9.

The MCDIFFaX fits of ice I$_{\text{sd}}$ from ASW and LDA from ice VIII have already been published in ref. 42\textcolor{white}{\cite{shephard2016LDAVIII}} and 97.\textcolor{white}{\cite{RosuFinsenThermal2022}} All other X-ray data has not been previously published.

\section{Acknowledgements}
We thank S. J. Cox and V. Kapil for stimulating discussions and suggestions, M. Vickers for help with the X-ray diffraction measurements and J. K. Cockcroft for access to a Cryojet HT.

\textbf{Funding}
We are grateful to the Materials Chemistry Consortium (Grant EP/L000202) and the UK Materials and Molecular Modelling Hub (Grants EP/P020194/1 and EP/T022213/1) for access to the ARCHER, Thomas, and Young supercomputers. We acknowledge funding from the European Research Council (ERC) under the European Union’s Horizon 2020 research innovation programme grant 725271 and the ``n-aqua'' ERC project (grant 101071937)).

\textbf{Competing interests} Authors declare that they have no competing interests.

\textbf{Author contributions}
\hfill \break
Conceptualization: MBD, CGS, AM.\\
Investigation: MBD, ARF, CGS, AM.\\
Funding acquisition: CGS, AM.\\
Software: MBD, CGS.\\
Visualisation: MBD, CGS.\\
Writing -- original draft: MBD.\\
Writing -- review \& editing: MBD, ARF, CGS, AM.\\


\bibliographystyle{ieeetr}
\bibliography{main}

\clearpage



\section*{Supplementary material}

**Supporting Materials for Low-Density Amorphous Ice Contains Crystalline Ice**

Michael Benedict Davies,[1,2] Alexander Rosu-Finsen,[3] Christoph G. Salzmann,[3, a)] and Angelos Michaelides[2, 1, b)]

[1)] *Department of Physics and Astronomy, University College London, London WC1E 6BT, UK*

[2)] *Yusuf Hamied Department of Chemistry, University of Cambridge, Lensfield Road, Cambridge CB2 1EW, UK*

[3)] *Department of Chemistry, University College London, 20 Gordon Street, London WC1H 0AJ, UK*

[a)] Electronic mail: c.salzmann@ucl.ac.uk

[b)] Electronic mail: am452@cam.ac.uk

## A.  Finite size effect when quenching liquid water

Fig. S3. compares the results of quenching a large system with 192,000 mW molecules (as presented in the main manuscript), to a smaller 3,000 mW molecule system. A clear difference in behaviour is observed which can be explained by the nucleation rates dependence on volume. This volume dependence means that for the smaller system such an optimal match to LDA is not found. The system volume is too small to statistically see multiple local homogeneous nucleation events. Thus the LDA structure with the optimal balance of heterogeneity at small scales (due to the presence of crystalline grains) but isotropy at large scales (due to the random orientation, location, and size of the grains) is not achieved.

## B.  Quenching TIP4P/2005

The nucleation rates of different water models allows different amorphous structures to be probed via quenching. The fully atomistic TIP4P/2005 water model[1] has a much slower nucleation rate and dynamics than mW. This enabled more amorphous states to be obtained when the same quenching procedure was carried out as illustrated in Fig. S4. Similar amorphous states have been reported in the literature.[2–4] Although these more amorphous structures give a reasonable agreement with experiment, it is much poorer than that of the partially crystallised states obtained by the mW model as can be seen in Fig. S4c. Moreover, the agreement follows the same trend as mW by improving as the proportion of crystallinity increases from 1.9 % to 4.1 % to 5.7 % to 8.1 %. As with mW, it appears a partially crystallised TIP4P/2005 system would give a very good agreement with experiment as well. However, the computational cost of obtaining higher levels of crystallinity for TIP4P/2005 is not practical in this study due to the model's order of magnitude increase in cost combined with the $\kappa$ required (on the order of 0.01 K ns$^{-1}$). In conclusion, quenching TIP4P/2005 has given further evidence that the LDA sample in experiment is not a fully amorphous state but instead a partially crystallized state.

## C.  Experiments on obtaining $I_{sd}$ from LDA.

The parent phase of LDA is either a different type of ice or the LDA has been prepared through different routes. The LDA is then heated to obtain ice $I_{sd}$ while XRD patterns are

recorded. The heating rates are the same for all data sets. As seen in Table S1, LDA as well as ice IX are formed when heating ice VIII. However, the amount of ice IX formed has previously been estimated to be at most 5 %,[5] and therefore deemed to have a minor effect on the resultant ice $I_{sd}$ diffraction pattern upon further heating.

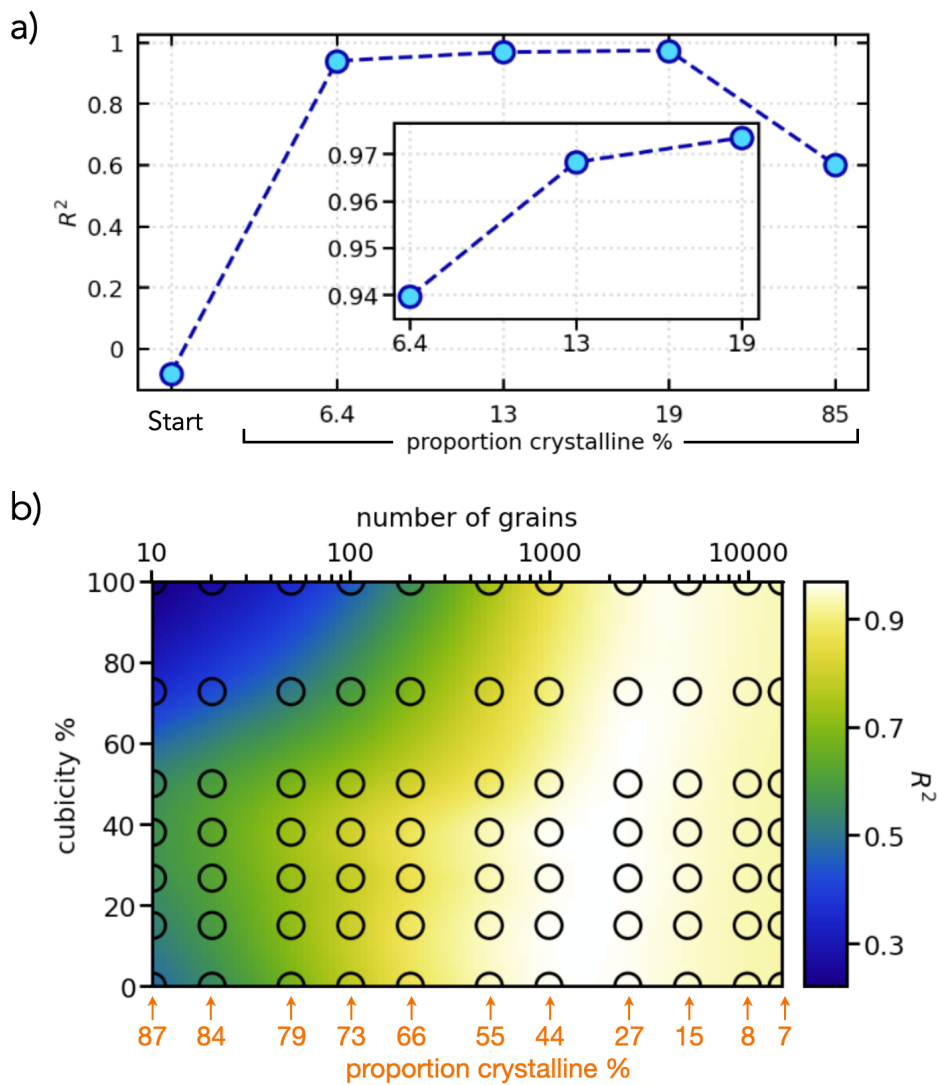

Supplementary Figure S1. Agreement between $S(Q)$ of computational models and experiment for LDA as quantified by the coefficient of determination ($R^2$). A negative $R^2$ is achieved if a fit is worse than a horizontal line at the mean of the data. (a) Structural models from quenching liquid water. (b) Structural models from building polycrystalline ice. Please note that $R^2$ is not the square of the Pendry R factor.

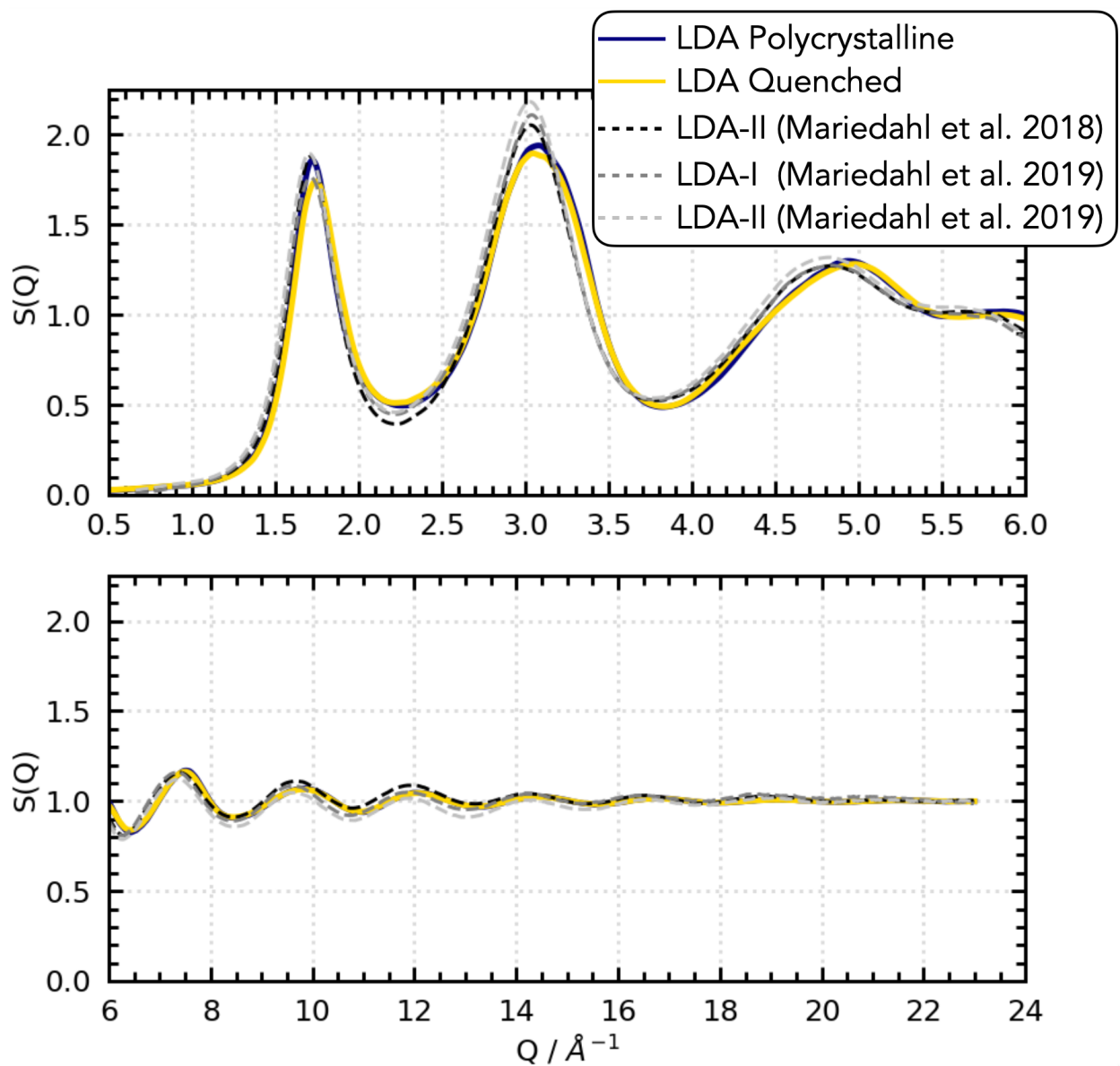

Supplementary Figure S2. Experimental $S(Q)$ of LDA-I and LDA-II measured by Mariedahl et al. (2019)[6] are plotted with the experimental LDA-II pattern of Mariedahl et al. (2018)[7] used in the main manuscript and the two optimal computational LDA models discovered in this study.

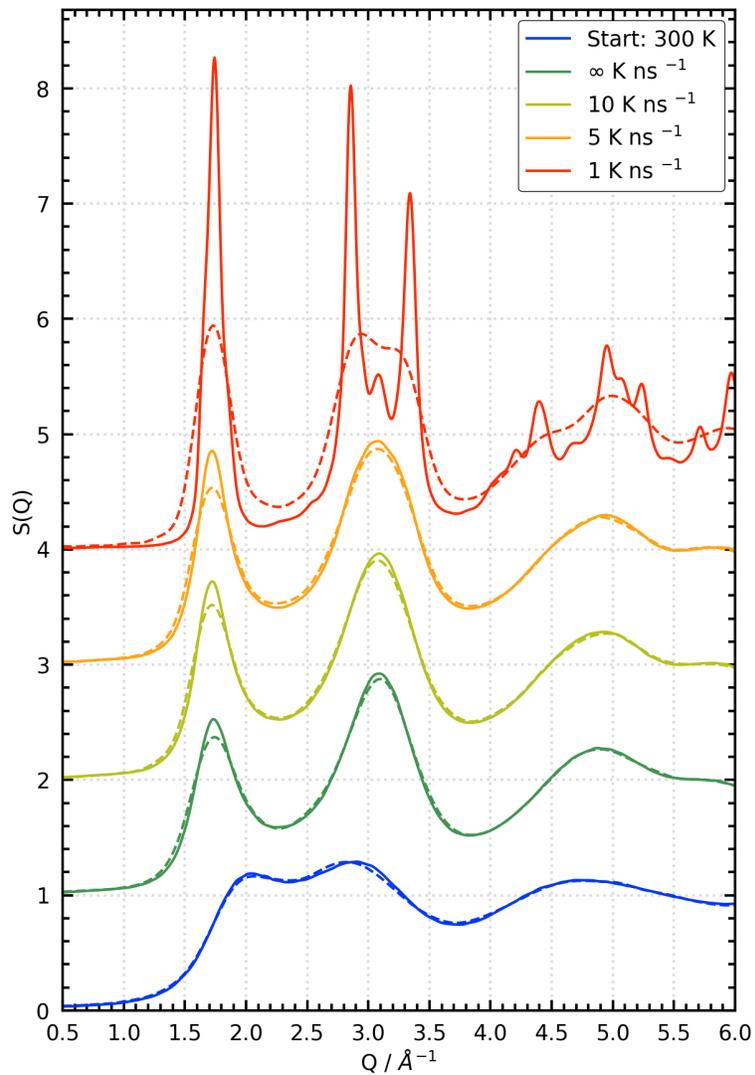

Supplementary Figure S3. A finite size effect when quenching liquid water in simulation. $S(Q)$ of approximately $21 \times 21 \times 21$ nm periodic simulation box with 192,000 mW molecules (solid lines) versus approximately $4 \times 4 \times 4$ nm periodic simulation box with 3,000 mW molecules (dashed lines).

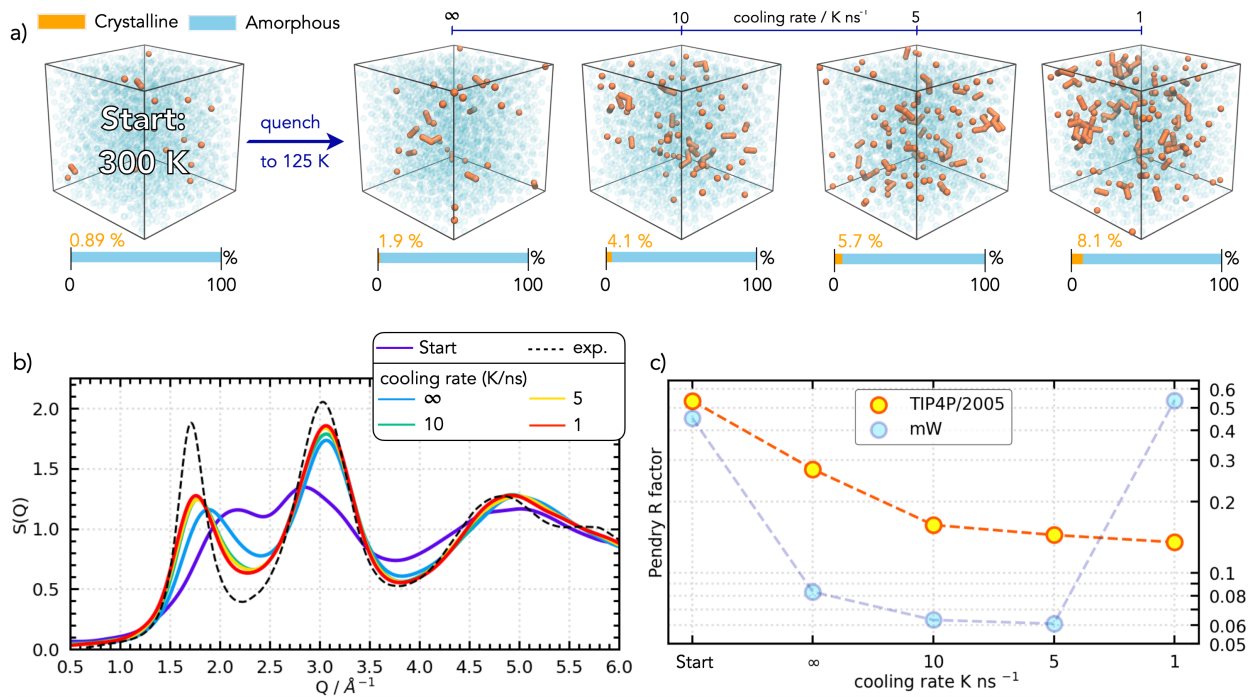

Supplementary Figure S4. Quenching liquid water using the TIP4P/2005 water model. (a) Snapshots of systems taken from quenching water from 300 K to 125 K at different cooling rates. (b) $S(Q)$ of quenched systems along with experiment (black dashed line). (c) Pendry R factor against cooling rate for TIP4P and mW water models.

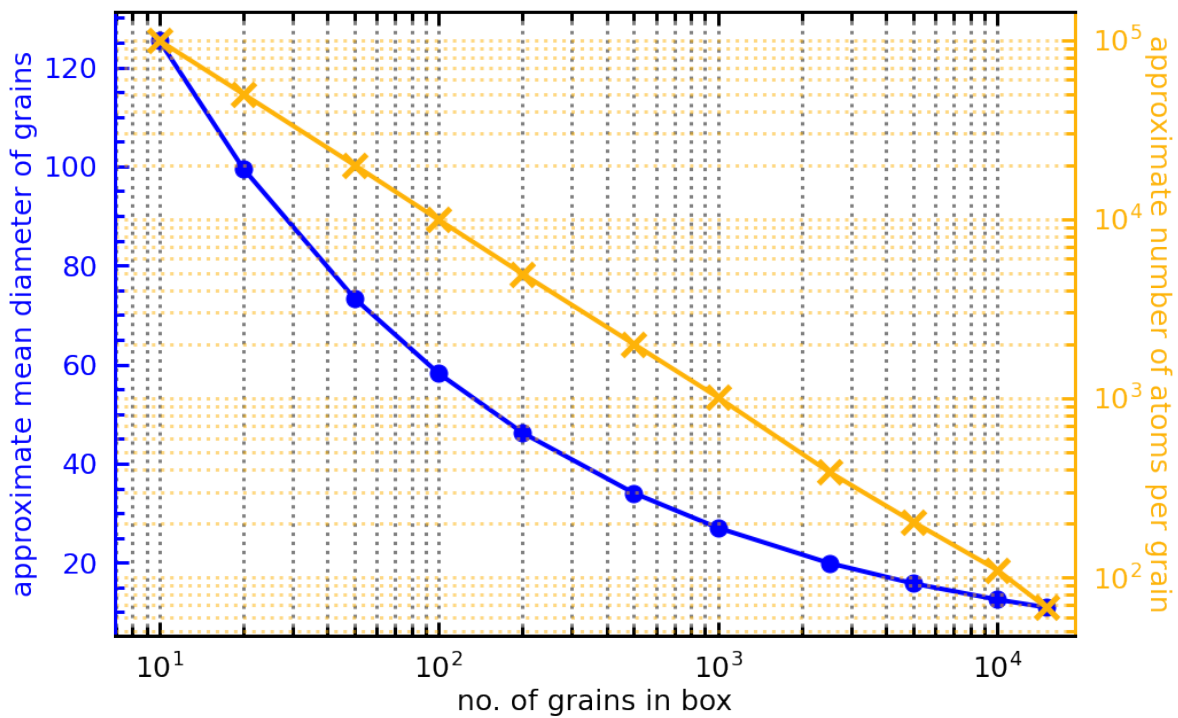

Supplementary Figure S5. Approximate mean diameter of ice grains (blue) and the approximate number of atoms per grain (orange) from the computational polycrystalline systems. Diameters are calculated by assuming that the grains are spherical. The number of atoms per grain is taken by cutting a sphere with the corresponding diameter from an ice $I_h$ structure at 0 K.

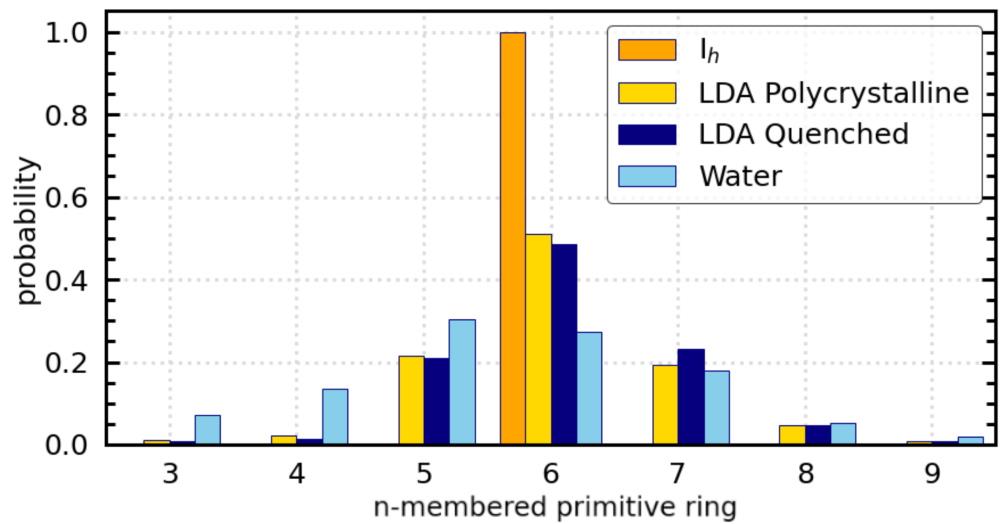

Supplementary Figure S6. Primitive ring distributions of liquid water, ice $I_h$, and the LDA models obtained in simulation.

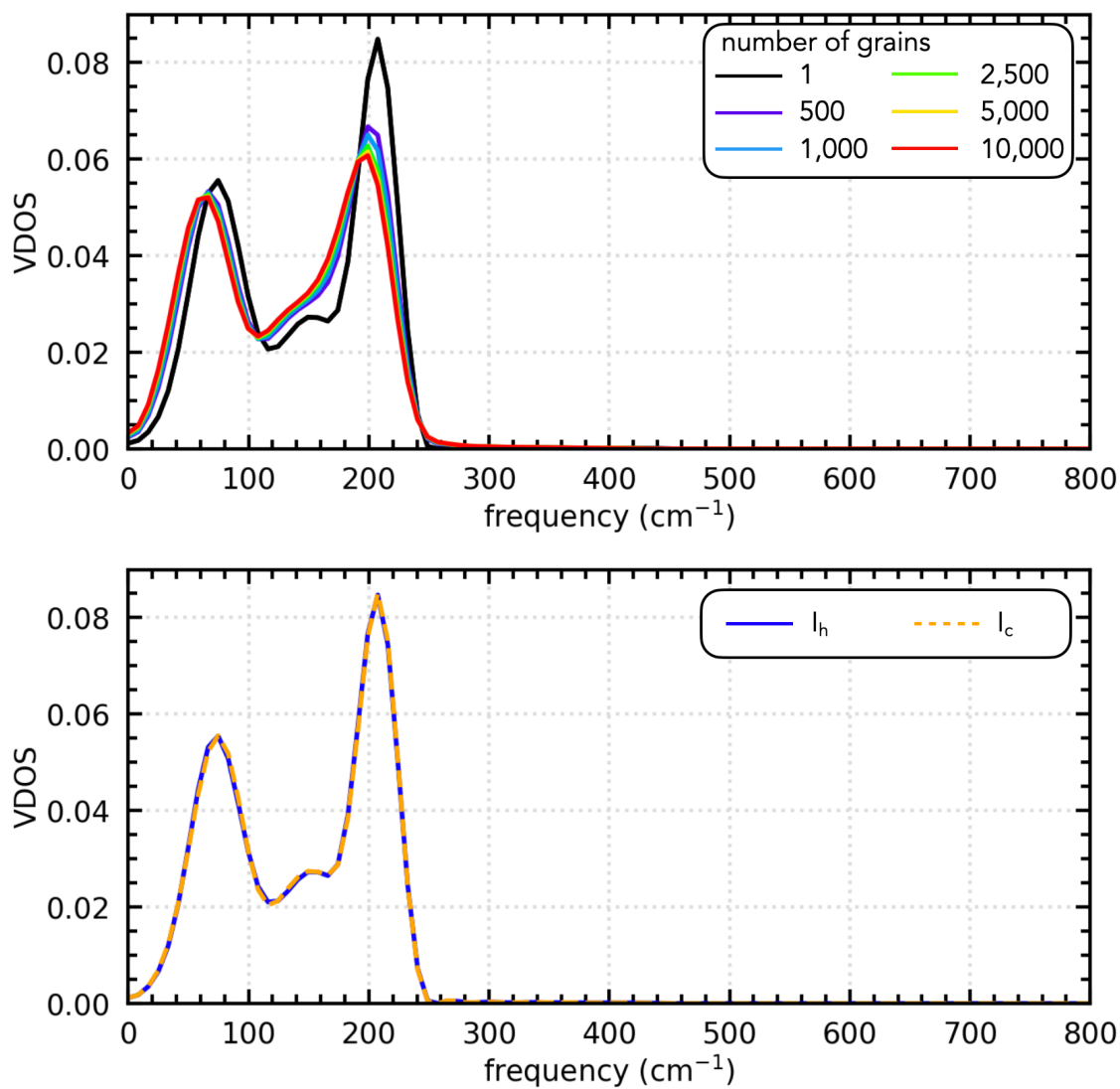

Supplementary Figure S7. Comparison of vibrational density of states (VDOS) for bulk and polycrystalline computational systems. (top) VDOS for different levels of polycrystallinity of ice at 38 % cubicity where the best match to LDA in experiment was found via $S(Q)$; one grain refers to the bulk structure. (bottom) VDOS for bulk ice $I_h$ and ice $I_c$.

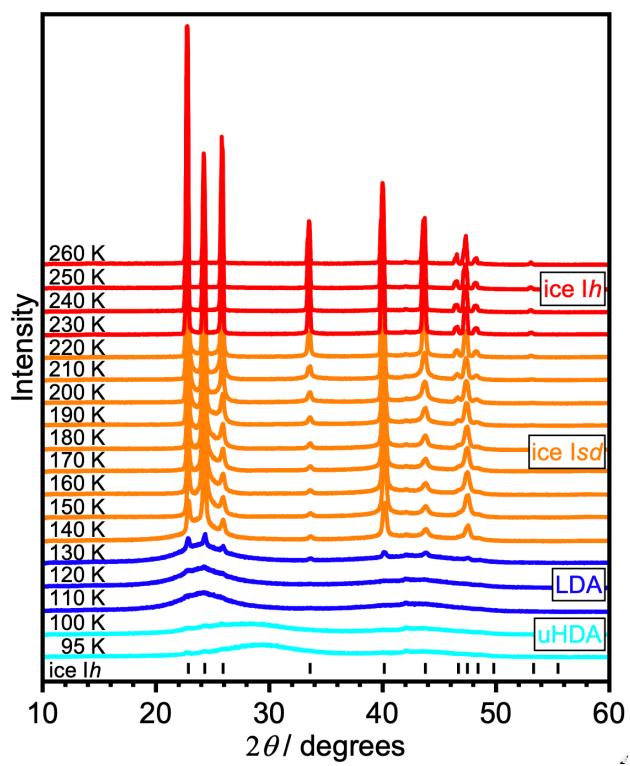

Supplementary Figure S8. Entire progression upon heating uHDA (cyan) in experiment, which transforms to LDA (blue) around 110 K, to ice $I_{sd}$ near 140-150 K and finally to ice $I_h$ near 220-230 K.

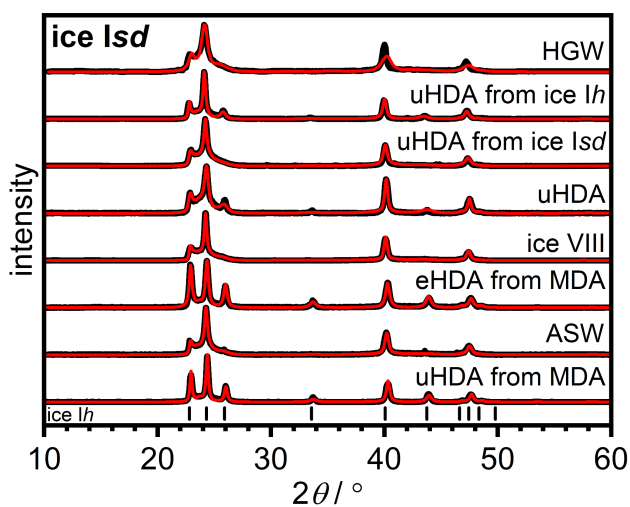

Supplementary Figure S9. Experimental ice $I_{sd}$ X-ray diffraction patterns immediately after the transition from LDA (black) and the corresponding MCDIFFaX fits (red).

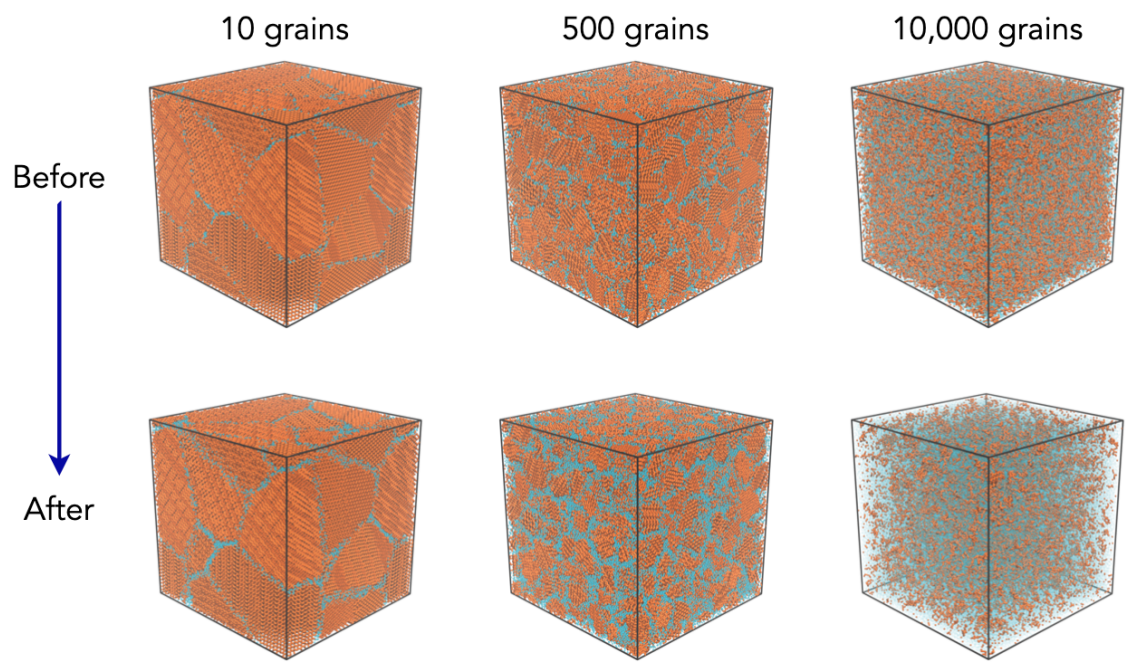

Supplementary Figure S10. Snapshots of the structures of the computational polycrystalline systems before and after annealing.

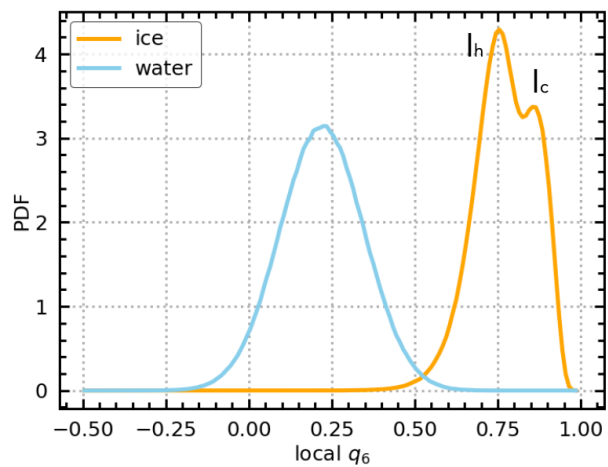

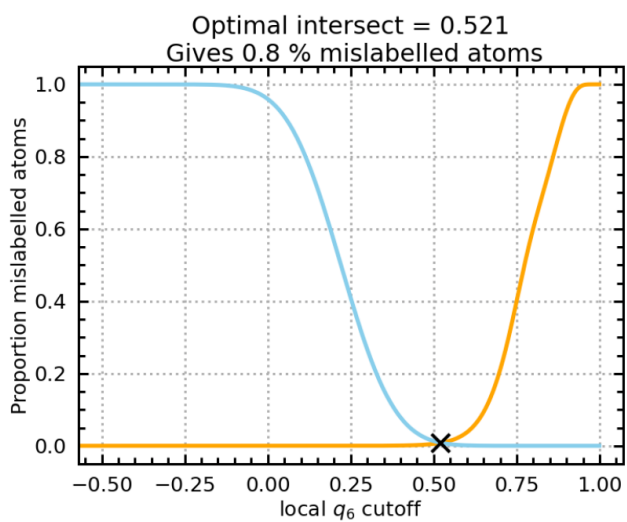

Supplementary Figure S11. Procedure to identify molecules with crystalline and amorphous local environments using the mW water model. (top) Distributions of local $q_6$ for water and ice. Distribution for ice comes from combining local $q_6$ from the bulk ice $I_h$ and ice $I_c$ simulations. (bottom) Proportion of mislabelled atoms for different choices of the local $q_6$ cutoff between crystalline and amorphous environments.

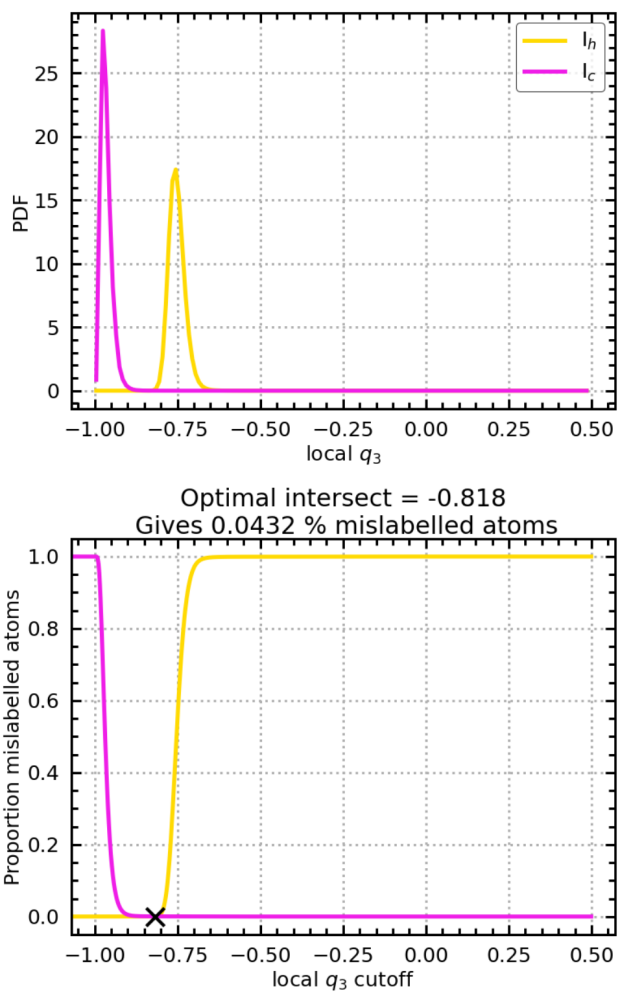

Supplementary Figure S12. Procedure to identify molecules with cubic and hexagonal local environments for the mW water model. (top) Distributions of local $q_3$ for ice $I_h$ and ice $I_c$. (bottom) Proportion of mislabelled atoms for different choices of the local $q_6$ cutoff between ice $I_h$ and ice $I_c$ environments.

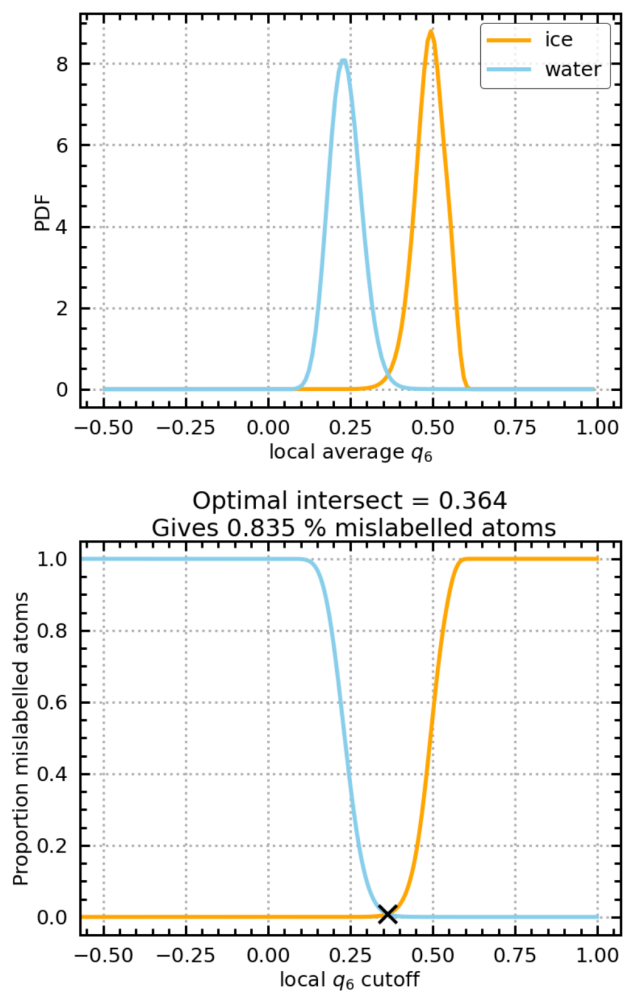

Supplementary Figure S13. Procedure to identify molecules with crystalline and amorphous local environments for the TIP4P/2005 water model. (top) Distributions of local $q_6$ for water and ice. Distribution for ice comes from combining local $q_6$ from the bulk ice $I_h$ and ice $I_c$ simulations. (bottom) Proportion of mislabelled atoms for different choices of the local $q_6$ cutoff between crystalline and amorphous environments.

TABLE S1. Transition of phases for converting LDA samples to ice $I_{sd}$ in experiment. (∗) Both ice $I_h$ and ice $I_{sd}$ were made in the press by first making ice II, removing the force to 100 N, and heating to beyond the transitions to ice $I_h$ or ice $I_{sd}$ before quenching and recompression to uHDA.

| Parent ice | Intermediate ice upon heating | Final product upon heating |
| --- | --- | --- |
| MDA compressed to uHDA | LDA | ice $I_{sd}$ and then ice $I_h$ |
| uHDA | LDA | ice $I_{sd}$ and then ice $I_h$ |
| ice $I_h$ compressed to uHDA* | LDA | ice $I_{sd}$ and then ice $I_h$ |
| ice $I_{sd}$ compressed to uHDA* | LDA | ice $I_{sd}$ and then ice $I_h$ |
| MDA compressed to eHDA | LDA | ice $I_{sd}$ and then ice $I_h$ |
| Ice VIII | LDA & ice IX | ice $I_{sd}$ and then ice $I_h$ |
| ASW/LDA | n/a | ice $I_{sd}$ and then ice $I_h$ |
| HGW/LDA | n/a | ice $I_{sd}$ and then ice $I_h$ |



\end{document}